\documentclass{mem}
\usepackage{natbib}\usepackage{txfonts}\usepackage{balance}
\usepackage{flushend}
\usepackage{graphicx}
\usepackage[breaklinks,pdftex]{hyperref}
\idline{75}{282}
\begin{document}
\def\teff{$T\rm_{eff }$}
\def\kms{$\mathrm {km s}^{-1}$}

\title{
Molecular complexity of young solar analogues}


\author{
E. \,Bianchi\inst{1} 
\and  M. \, De Simone\inst{2, 3} \and  G. \, Sabatini\inst{3} \and J. \, Frediani\inst{4,2} \and L. \, Podio\inst{3} \and C. \, Codella\inst{3}  
\institute{Excellence Cluster ORIGINS, Boltzmannstraße 2, 85748, Garching bei Mu\"unchen, -- Germany
\and
 European Southern Observatory, Karl-Schwarzschild-Straße 2 D85748 Garching bei M\"unchen, Germany
  \and 
 INAF, Osservatorio Astrofisico di Arcetri, Largo E. Fermi 5, I-50125, Firenze, Italy
 \and 
 AlbaNova University Centre, Department of Astronomy, Stockholm University, SE-10691, Sweden}\\
\email{eleonora.bianchi@origins-cluster.de}
}

\authorrunning{Bianchi et al.}

\titlerunning{Molecular complexity of young Solar analogues}

\date{Received: XX-XX-XXXX (Day-Month-Year); Accepted: XX-XX-XXXX (Day-Month-Year)}

\abstract{
How does molecular complexity emerge and evolve during the process leading to the formation of a planetary system? 
Astrochemistry is experiencing a golden age, marked by significant advancements in the observation and understanding of the chemical processes occurring in the inner regions of protostellar systems. However, many questions remain open, such as the origin of the chemical diversity observed in the early evolutionary stages, which may influence the chemical composition of the forming planets. Additionally, astrochemistry provides us with powerful tools to investigate the accretion/ejection processes occurring in the inner regions of young embedded objects, such as jets, winds, accretion streamers, and shocks.
In this chapter, we review the observational efforts carried out in recent years to chemically characterize the inner regions of Solar-System analogs. We summarize our current understanding of molecular complexity in planet-forming disks and shed light on the existing limitations and unanswered questions. Finally, we highlight the important role of future radio facilities, like SKAO and ngVLA, in exploring the chemical complexity of the regions where planetary systems are emerging.
}
\maketitle{}

\section{Introduction}\label{sec:intro}
\begin{figure*}[t!]
\includegraphics[scale=0.33]{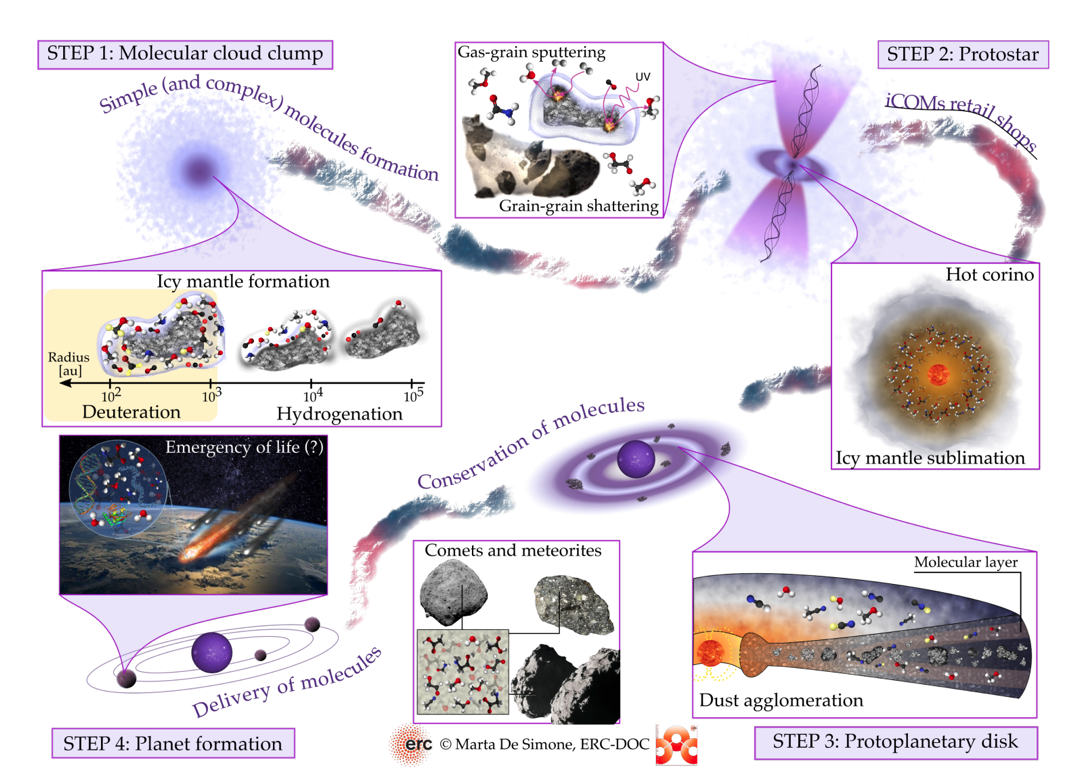}
\caption{\footnotesize
Representation of the chemical evolution during the early stages of a Solar-like star formation (adapted from \citealt{de_simone_phdthesis_2022}). 
}
\label{fig:sketch-astrochemistry}
\vspace{-0.5cm}
\end{figure*}
The emergency of Life on our planet is the result of a complex interplay of physical and chemical processes. Simple and complex interstellar molecules are observed in all the different formation stages of a Sun-like star, from the early prestellar stage (age $<$10$^4$ yr from the beginning of the gravitational collapse), to the protostellar (Class 0, $\sim$10$^{4}$ yr and Class I, $\sim$10$^{5}$ yr) stages as well as in comets and meteorites, representative of the pristine material from which our Solar System formed \citep[e.g., see the review by][]{Ceccarelli2023}.
The sketch in Fig. \ref{fig:sketch-astrochemistry} illustrates how molecules are formed on the icy mantles of dust grains and in the gas-phase during the cold evolutionary stages and then transferred to the later stages and incorporated into planets. Of particular interest in the context of the origin of Life are the so called interstellar Complex Organic Molecules (iCOMs, \citealt{Herbst-vanDishoeck09,Ceccarelli2023}). These species contain more than 6 atoms and heteroatoms (such as C, O, N). iCOMs constitute the bricks to form more complex and prebiotic species. They are observed at all stages but, more abundantly in the inner protostellar regions ($<$ 100 au), where the gas temperature is so high to induce the thermal sublimation of the species frozen-out in the icy grain mantles, so to trigger a complex chemistry. This dense ($>10^7$ cm$^{-3}$) and hot ($>$ 100 K) region rich in iCOMs is called hot corino \citep{ceccarelli_hot_2004}.

However, not every protostar possesses a hot corino, and significant differences have been observed in the millimeter (mm) spectra of protostars. Indeed, besides hot corinos (rich in iCOMs), there are the so-called Warm Carbon Chain Chemistry (WCCC) sources, poor in iCOMs and rich in carbon chains and rings \citep{sakai_warm_2013}.
The origin of this chemical diversity may be related to a different collapse timescale which lead to a different composition of the icy dust grain mantles, or to external environmental effects \citep[e.g.,][]{aikawa_chemical_2020}. In addition, complex carbon species (with at least seven C atoms) are challenging to observe at (sub-)mm wavelengths but have been found to be abundant in few cold prestellar cores using radio observations \citep{Bianchi2023a, McGuire2020, Cernicharo2021}.
Their study adds an important piece of the overall puzzle as they might have a crucial role in the heritage of organic material from the pre- and proto-stellar phases to the objects of the newly formed planetary system, like comets (e.g. \citealt{Mumma2011}). 

Furthermore, 
several recent evidences suggest that planetesimal formation may start during the early stages of their formation \citep[$\lesssim$ 0.5 Myr; e.g.,][]{Manara2018,Tychoniec2020}. In this respect, the exploration of the chemical content of young protostellar disk is key to constrain the initial conditions of the planet formation process. 
Additionally, young disks are also characterised by accretion/ejection activity, such as large-scale outflows and accretion streamers, i.e. gas and dust infalling from the outer envelope regions to the disk \citep[e.g.,][]{Pineda2023}.
These processes result in a complex interplay between the protostars and their surrounding medium which mutually modifies the initial physical-chemical properties. The supply of energy and momentum leads to feedback mechanisms such as stellar winds, jets and outflows, which in turn create a dynamic environment that paves the way for the iCOMs formation \citep[e.g.,][]{Codella17, de_simone_seeds_2020}. Accretion streamers also deliver mass from the outer envelope regions to the disk at a rate larger than the mass accretion rate from the disk to the protostar, with a huge impact on the mass available to form planets, and on the disk stability \citep[e.g., ][]{Pineda2023,flores23}. They also produce accretion shocks when impacting onto the disk, thus sputtering the ices in the disk upper layers and altering their chemical composition \citep[e.g., ][]
{maureira2022,
bianchi_2022,Bianchi2023a}.

We here present some results from the Fifty AU STudy of the chemistry in the disk/envelope system of Solar-like protostars (FAUST\footnote{http://faust-alma.riken.jp/index.html}; \citealt{Codella2021}) ALMA\footnote{https://www.almaobservatory.org/en/home/} large program. 
We also discuss the limitations of (sub-)mm observations and the perspectives for new generation of radio interferometers.

\section{The FAUST ALMA Large Program}\label{sec:faust}

FAUST \citep{Codella2021} is the first ALMA large program entirely dedicated to the chemical exploration Solar-System analogs. More specifically, it focuses on 13 nearby (137–235 pc) star forming regions associated with multiple young low-mass protostellar systems (Class 0/I). This ambitious project has been designed to probe the entire range of relevant spatial scales, from the large-scale envelope ($\sim$ 2000 au) to the inner disk/jet system, imaged at the Solar System scale (down to 50 au). This is achieved thanks to a unique combination of different array configurations including the 12-m array and the 7-m compact array (ACA). Three different spectral setups have been used to image numerous molecular tracers including iCOMs (CH$_3$OH, NH$_2$CHO, CH$_3$CHO, CH$_3$OCH$_3$, and HCOOCH$_3$), shock tracers (SO, SiO), and deuterated species (c-C$_3$HD, N$_2$D$^+$, HDCO, D$_2$CO, and CH$_2$DOH). FAUST observations offers a powerful dataset which allows to combine different molecular tracers in order to discover and disentangle all the processes associated with Sun-like star formation. 

\section{Chemical complexity at the protostellar stage with FAUST}\label{sec:protostars}

\subsection{Chemical segregation in the IRAS4A2 hot-corino}\label{sec:jenny}
\begin{figure}
    \vspace{-0.3cm}\includegraphics[width=0.92\columnwidth]{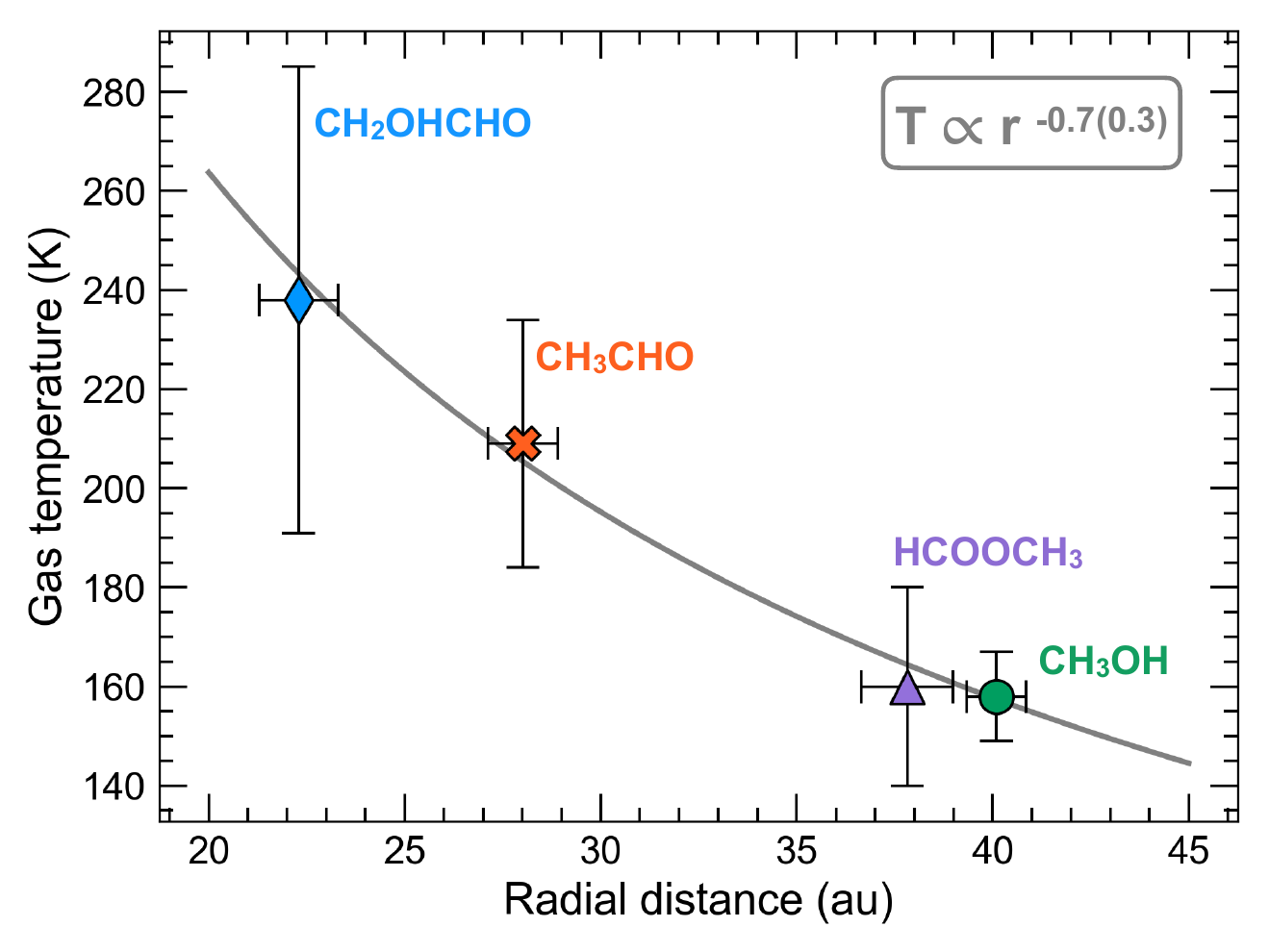}
    \caption{\footnotesize Rotational temperature vs emitting radius in the IRAS 4A2 hot corino. Each species is shown with different colors and markers. 
    The grey solid line indicates the fit performed on the data points.}
    \label{fig:jenny}
    \vspace{-0.5cm}
\end{figure}

\begin{figure*}[htp!]
\centering
\includegraphics[width=0.8\textwidth]{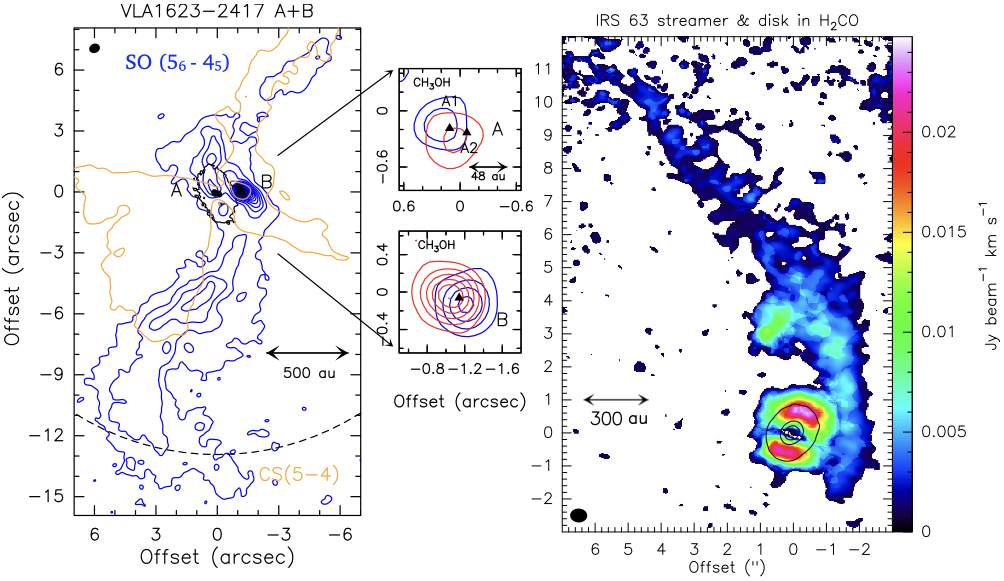}
    \caption{\footnotesize Accretion streamers feeding the disks of 
    the VLA 1623-2417 protostellar system ({\it Left panel}) and the Class I disk IRS~63 ({\it Right panel}).
Dust continuum is in black contours.
{\it Left panel}: CS(5--4) (orange) shows the outflow cavities, SO(5$_{\rm 6}$--4$_{\rm 5}$) (blue) an extended accretion streamers feeding the protostars from the south; CH$_3$OH(5$_{\rm 1,4}$--4$_{\rm 1,3}$ A) the molecular enrichment around the rotating protostellar disks in A and B \citep{Ohashi2022,Codella2022,Codella2024}.
{\it Right panel}: Integrated intensity (moment 0) map of H$_2$CO $3(0,3)-2(0,2)$ towards IRS 63 revealing the gaseous disk and the streamer feeding it (Podio et al. subm.).}
    \label{fig:lindaclaudio}
\vspace{-0.4cm}
\end{figure*}

We used the emission of iCOMs as probes of the structure of the inner regions of IRAS 4A2, a well-studied hot corino \citep[e.g.,][]{lopez_2017, marta_2017,de_simone_hot_2020}. Exploiting different species at high spatial resolution ($\sim$50 au), we resolved for the first time its emitting size, and attempted to measure the gas temperature as a function of the distance from the central protostar (Frediani et al. in prep.). 
We imaged IRAS 4A2 in 4 iCOMs: CH$_3$OH, HCOOCH$_3$, 
CH$_3$CHO, and CH$_2$OHCHO, 
and derived rotational temperatures and column densities mainly via a Local Thermodynamic Equilibrium (LTE) radiative transfer analysis. By performing an image-plane modelling of the source, we found strong evidence of spatial segregation between different species. For the first time we resolved the IRAS 4A2 hot corino that shows a onion-like structure with CH$_3$OH tracing the most extended gas, followed by CH$_3$CHO and HCOOCH$_3$, up to the most compact and unresolved CH$_2$OHCHO.  
Assuming that the rotational temperature is a good approximation for the gas temperature, we attempted to retrieve the temperature profile of the Class 0 IRAS 4A2.
The preliminary results suggest a trend between the emitting radius and the measured temperature, consistent with a disk-like power-law decay ($\beta \approx$ 0.7) with the radial distance from the protostar (Fig. \ref{fig:jenny}). 
As conclusive remarks, we highlight the crucial role of high angular resolution observations ($<$50 au) combined with a multi-line analysis to retrieve the physical properties of hot corinos in an unprecedented way. 

\subsection{Accretion streamers feeding disks} \label{sec:linda}

We investigated the effect of the interaction with the environment on the physical and chemical structure of young disks by studying with ALMA-FAUST the disks-streamers interaction towards the Class 0 triple system VLA 1623-2417, and the Class I disk IRS~63, located in Ophiuchus.
Fig. \ref{fig:lindaclaudio} (Left) shows the different components of the triple protostellar system VLA 1623-2417 (A1, A2, and B) where: (i) CS traces the molecular bipolar cavity opened by protostellar jets, while (ii) CH$_3$OH shows the counter rotating disks around A1 and B; and (iii) SO reveals an extended accretion streamer feeding the protostars at a rate of $\sim$ 3--5 $\times$ 10$^{-7}$ $M_{\rm \sun}$ yr$^{-1}$ \citep{Ohashi2022,Codella2022,Codella2024}.
The latter is comparable to the mass accretion rate of VLA1623 B, stressing the importance of streamers in contributing to the final mass of protostellar disks.


Fig. \ref{fig:lindaclaudio} (Right) shows H$_2$CO towards IRS~63 which traces the disk and the streamer, first imaged in C$^{18}$O by \citet{flores23}.
The streamer connects the disk with the outer regions of the envelope out to $\sim$1500 au. 
In Podio et al. (submitted) we reported for the first time the detection of the H$_2$CO deuterated isotopologues, HDCO and D$_2$CO, both in the disk and in the streamer that is feeding it, thus allowing us to estimate the deuterium fraction of H$_2$CO. 
The D$_2$CO intensity as well as the estimated deuterium fraction, are highly asymmetric in the disk and peak where the large scale H$_2$CO streamer impacts onto the disk (i.e. the SE disk side). Deuteration in the streamer is larger or comparable to that measured in the disk, indicating that the streamer delivers deuterated material from the cold outer regions of the envelope to the disk, thus deeply impacting its chemical composition.


\subsection{Protostellar Feedback in Cra}\label{sec:giovanni}

\begin{figure}
\centering
    \includegraphics[width=1\columnwidth]{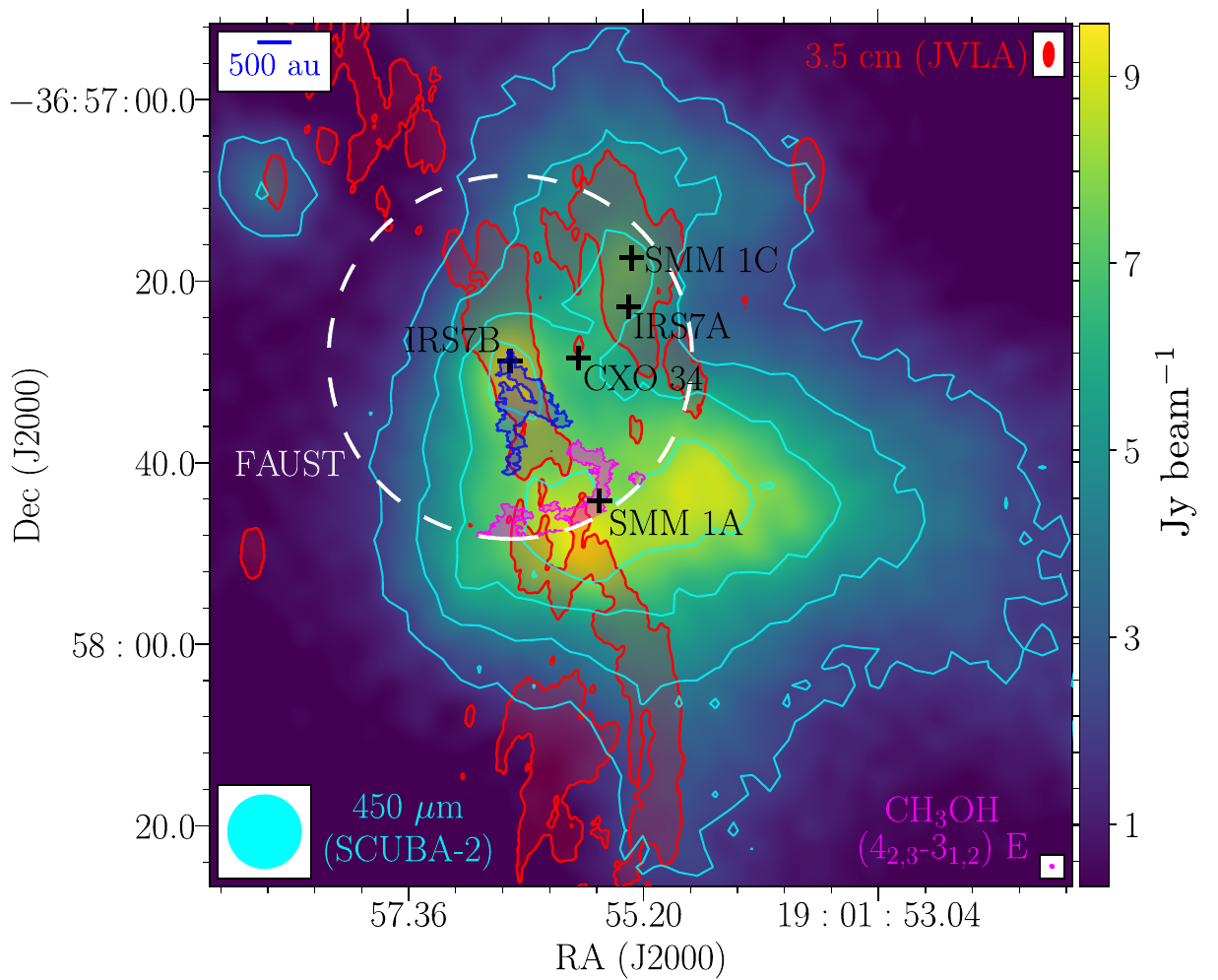}
\caption{\footnotesize 
SCUBA-2 map at 450 $\mu$m (color map and cyan contours) of the CrA cluser superimposed with JVLA 3.5 cm map \citep[red contours;][]{Liu14}, and CH$_3$OH (4$_{2,3}$-3$_{1,2}$)~E (magenta contours) and dust continuum emission at 1.3 mm (blue-shaded area) observed in FAUST \citep[see][]{Sabatini24}. 
Beams are reported in the plot corners.
}
    \label{fig:sabba}
    \vspace{-0.5cm}
\end{figure}

Among the FAUST targets, the nearby Corona Australis (CrA) region is associated with several Class 0/I protostellar systems \citep[e.g.][]{Knacke73}.
IRS7B, located in the heart of the CrA cluster, has been resolved into a binary system, with each component associated with discs aligned at a PA of 115$^\circ$ and an inclination of $\sim$ 65$^\circ$ \citep{Takakuwa24}. The cluster also harbours CXO 34, IRS7A, and SSM~1C, two Class I and one Class 0 protostars, respectively. These studies also revealed significant continuum emission from SMM~1A, observed with SCUBA-2 at 450$\mu$m (Fig.~\ref{fig:sabba}). However, the exact nature of this emission is uncertain, and interpreted either as a quiescent prestellar object \citep{Nutter05} or as a filamentary structure externally illuminated by the Herbig Ae/Be star R-CrA \citep[e.g.][]{LindbergJorgensen12}.

We present in Figure \ref{fig:sabba} high-resolution ($\sim$50 au) maps of 1.3 mm dust continuum, SiO, CH$_3$OH and H$_2$CO towards CrA. Methanol shows an arc-like structure at $\sim$1800~au from IRS7B (magenta in Fig.~\ref{fig:sabba}), close to SMM 1A, and aligned perpendicular to the disc major axis. The arc is at the edge of two elongated structures detected in continuum at 1.3 mm (deep blue in Fig.~\ref{fig:sabba}), forming a cone emanating from IRS7B. 
These data reveal for the first time a bow shock driven by IRS7B and a two-sided dust cavity created by the mass-loss process. 
These results provide the first evidence for a symmetric dusty cavity opened by the jet of a low-mass protostar \citep{Sabatini24}.

\section{Living ALMA, preparing SKAO}\label{sec:radio}

The presented results show that observations at (sub-)mm wavelengths and, in particular, with ALMA interferometer are a very powerful tool to characterize the chemical and dynamical complexity of Solar-like protostars. Nevertheless, they have some limitations which hamper to build a comprehensive understanding. The most important is related to the high dust opacity which completely obscure molecular emission in young sources (Fig. \ref{fig:icemantles}) and, especially in the disk midplane \citep{Lee2017}. In addition, complex carbon species can not be studied at (sub-)mm wavelengths having their emission peak at longer (radio) wavelengths (see Fig. \ref{fig:c6h}; \citealt{Bianchi2023a}). The new generation of radio telescopes, such as the Square Kilometer Array Observatory (SKAO) and the next-generation VLA (ngVLA), will represent a major step ahead. On one hand, they will allow us to observe the midplane disk region in a frequency range where dust is optically thin. On the other hand, they will enhance our understanding of chemical complexity at early stages by enabling the observation of new complex molecular species. In the next section, we present some preliminary works done using observations at radio frequencies, offering new perspectives for exploring the chemistry of Solar-System analogs.

\subsection{Revealing the chemical history of young protostars at cm-wavelenghts}\label{sec:marta}

\begin{figure}
    \centering
\includegraphics[width=0.8\columnwidth]{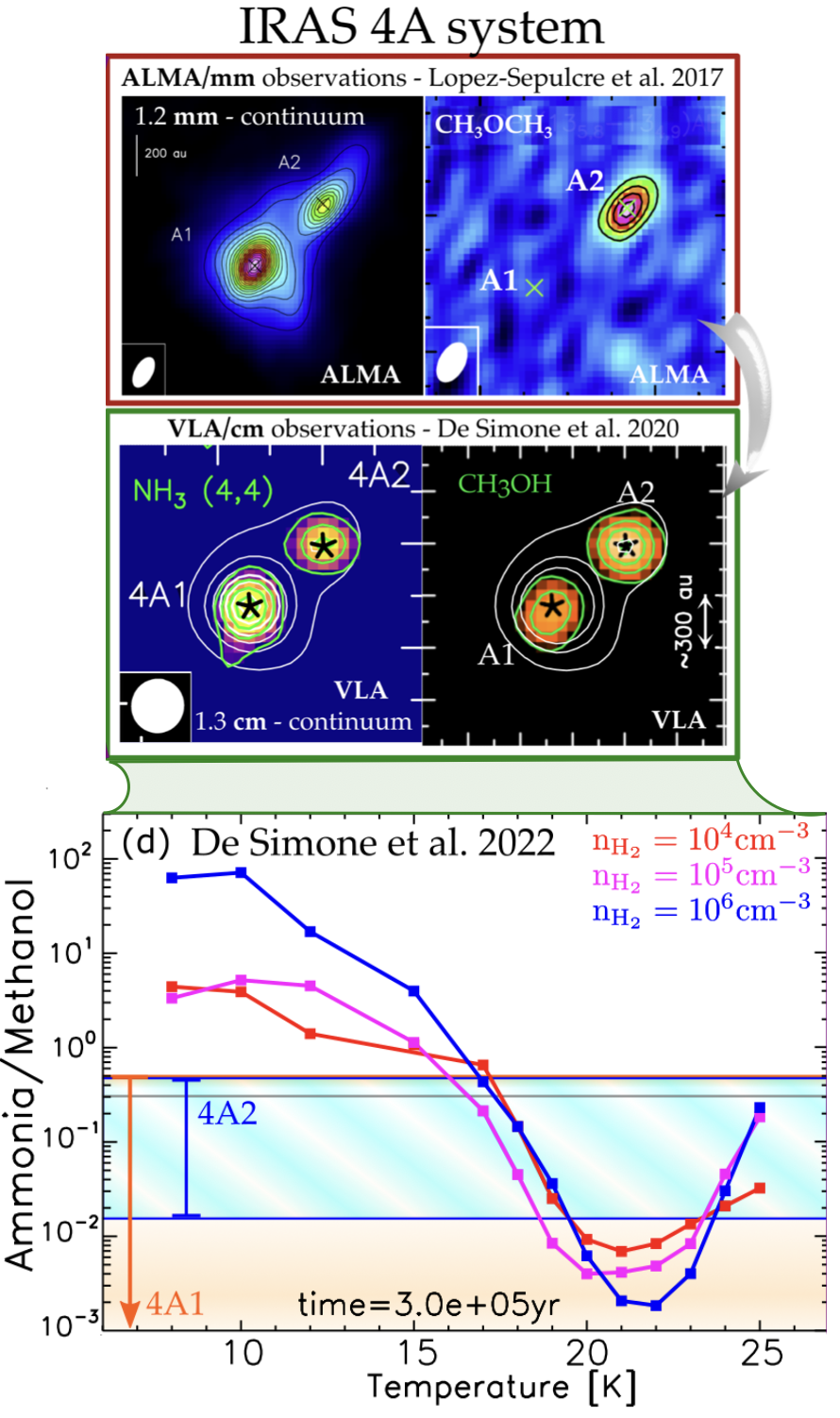}
    \caption{\footnotesize \textit{Upper panels}: IRAS 4A at mm (top) and cm (bottom) wavelengths. The mm dust is so thick to obscure the 4A1 hot corino that pops up in the cm with NH$_3$ and CH$_3$OH \citep{de_simone_hot_2020} 
    \textit{Lower panel}: NH$_3$/CH$_3$OH ratio versus dust temperature at different prestellar core densities after 0.3 Myr, assumed to be the ice mantles formation timescale \citep[other timescales are shown in][]{de_simone_tracking_2022}. The colored bands show the observed ratio in 4A2, 4A1.  
    }
    \label{fig:icemantles}
    \vspace{-0.5cm}
\end{figure}

We observed three protostars
(IRAS 4A2, 4A1, 4B) in NGC 1333 with the VLA at radio frequencies, to test if the chemical diversity observed in the protostellar stage is due to observational biases in the (sub-)mm \citep{de_simone_hot_2020} or to a different chemical composition of the dust grain mantles \citep{de_simone_tracking_2022}. Directly study the ice mantle composition with IR observations can be challenging for these embedded objects. The protostars have been imaged at 300 au scale ($\sim 1''$) with CH$_3$OH and NH$_3$, which are among the major ice mantle components \citep{McClure_ice_2023} and can be observed in the gas-phase once released by the hot corino activity.
Additionally, they have well known formation paths \citep{ rimola_combined_2014,le_gal_interstellar_2014, 
Tinacci_theoretical_2022}, and their abundance ratio on the icy mantles depends on the cloud physical properties 
\citep{taquet_multilayer_2012, aikawa_chemical_2020}.
We derived the NH$_3$/CH$_3$OH abundance ratio using a multi-line analysis with a non-LTE large velocity gradient (LVG) approximation \citep[using \texttt{grelvg;}][]{ceccarelli_theoretical_2003}. We found similar values for both protostars (Fig. \ref{fig:icemantles}) indicating a similar chemical history, i.e. they were formed from pre-collapse material with similar physical conditions.
Fig. \ref{fig:icemantles} shows the theoretical prediction of NH$_3$/CH$_3$OH \citep[done with the astrochemical \texttt{GRAINOBLE} model;][]{taquet_multilayer_2012} as a function of the dust temperature, the gas density and the timescale. Comparing the observations with the predictions, we concluded that the dust temperature at the moment of the ice mantle formation was $\geq$ 17 K. This is relatively warm for a typical prestellar core in the inner 100 au \citep[e.g., L1544, and L1498 have about 7 K at these scales, derived from different molecular tracers;][]{crapsi_observing_2007, lin_deuterium_2024}. 
This suggests that the protostars did not have the usual dense and cold precollapse phase, as their mantles were mostly built during a relatively warm phase, 
which is characteristic of the less dense cloud material in NGC 1333 south 
\citep{zhang_herschel_2022}.
In other words, something must have happened that suddenly compressed the gas and triggered a fast collapse and the protostars formation. A possibility is that the collapse have been brutally started by the clash of an external bubble with NGC 1333 \citep{dhabal_connecting_2019, de_simone_train_2022}. 
Finally, we highlight the crucial role of cm arcsecond observations, combined with mm. 
Indeed, we constrained the chemical and dynamical history of the IRAS 4 protostars without being biased by dust opacity effects.

\subsection{Detecting Heavy Carbon Species}\label{sec:claudio}

\begin{figure}
    \centering
    \includegraphics[width=0.55\columnwidth]{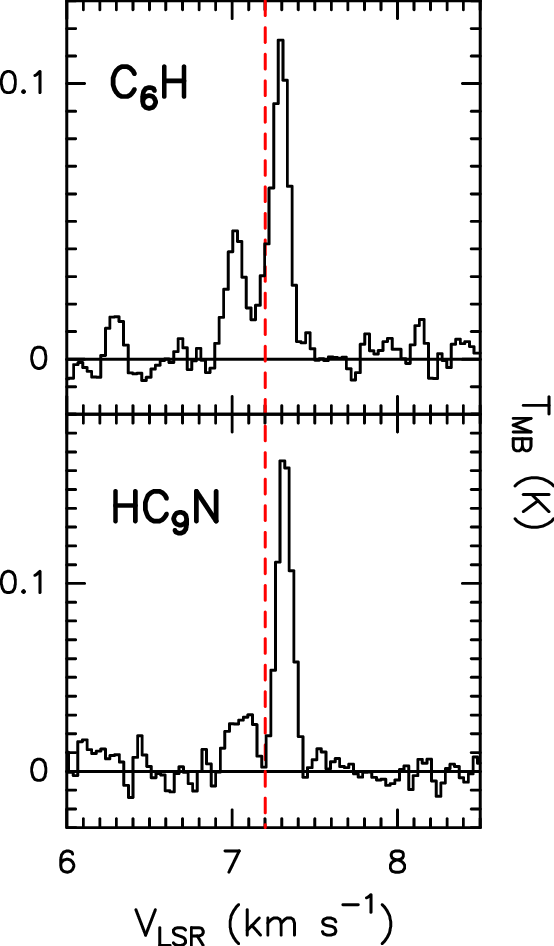}
    \caption{\footnotesize Example of C$_6$H ({\it Upper panel}) and HC$_9$N ({\it Lower panel}) line emission detected towards the prestellar core L1544 using the GBT antenna \citep{Bianchi2023a}. The dashed vertical line indicates the systemic source velocity.}
    \label{fig:c6h}
    \vspace{-0.5cm}
\end{figure}

Complex carbon chains and rings have their brightest transitions at frequencies lower than 50 GHz and can only be detected using radio telescopes. The main facilities so far available to this purpose are the 100 m GBT, the 40m YEBES antennas, and the VLA array interferometer.
We performed a spectral survey with the GBT at 8--15 GHz towards the prestellar core L1544. We detected a large number of emission lines due to long cyanopolyynes (HC$_7$N, HC$_9$N) and complex carbon species such as C$_3$S, C$_4$H, and C$_6$H \citep{Bianchi2023a}. The lines are spectrally resolved, with a double-peak profile (Fig. \ref{fig:c6h}), indicating a common emitting region.
More specifically, the emission peak is red-shifted in the southern part of the core, at the edge of the molecular cloud. We concluded that the formation of complex C-species is related to the abundance of free carbon atoms, produced by CO dissociation in the outer part of the core exposed to external UV-illumination.
A similar complex chemistry has been observed in the starless core TMC-1 by: (i) the GBT GOTHAM project \citep{McGuire2020} and (ii) the YEBES QUIJOTE survey \citep{Cernicharo2021}. A large number of heavy C-bearing species (as complex as e.g. c-C$_9$H$_8$, and o-C$_6$H$_4$) have been indeed discovered \citep[e.g.][and references therein]{
Siebert2022,
Cernicharo2021,Cernicharo2023,
Remijan2023}. 
Furthermore, complex carbon species have been observed in other 4 cores \citep{Burkhardt2021} suggesting a relevant and widespread chemical complexity so far mostly unexplored. Thanks to the new generation of radio facilities, such as the ngVLA and SKAO interferometers, it will be possible to reveal the complex carbon chemistry in small regions associated with the formation of Sun-like analogues and their planetary systems.

\section{Conclusions}\label{sec:conclusion}

Thanks to systematic surveys such as the FAUST project, we are advancing toward a global understanding of the physical-chemical processes around Solar-System analogs. We can now i) explore their chemical diversity down to spatial scales where the first planetesimals start to form ($\leq$ 50 au); ii) investigate the impact of feedback processes, such as outflows, winds, large-scale streamers and accretion shocks at disk scales; ii) explore the effects of global evolution of different environments on the chemical diversity of protostellar systems, shedding light on the composition of the resulting planetary systems.
A significant advancement in the coming years will be the advent of ngVLA and SKAO, enabling us to study the inner regions of Sun-like protostellar disks, including the midplane region, without biases from high dust opacity. Finally, they will enable us to uncover the reservoir of complex carbon species and take a significant step towards understanding the origins of Life.



\begin{acknowledgements}
CC, LP, and GS acknowledge 
the PRIN MUR 2020 BEYOND-2p 
(Prot. 2020AFB3FX),
the PRIN MUR 2022 FOSSILS 
(Prot. 2022JC2Y93), the project ASI-Astrobiologia 2023 MIGLIORA
(F83C23000800005), the INAF-GO 2023 PROTO-SKAO 
(C13C23000770005), the INAF Mini-Grant 2022 “Chemical Origins”, and the INAF-Minigrant 2023 TRIESTE.
JF acknowledges the DIFA, ESO Office for Science, and the Italian MIUR through the grant Progetti Premiali 2012 – iALMA (CUP C$52$I$13000140001$). EB acknowledges the Deutsche Forschungsgemeinschaft (DFG, German
Research Foundation) under Germany´s Excellence Strategy – EXC 2094 – 390783311. 
This paper makes use of the following ALMA data: ADS/JAO.ALMA\#2018.1.01205.L (PI: S. Yamamoto). ALMA is a partnership of the ESO (representing its member states), the NSF (USA) and NINS (Japan), together with the NRC (Canada) and the NSC and ASIAA (Taiwan), in cooperation with the Republic of Chile. The Joint ALMA Observatory is operated by the ESO, the AUI/NRAO, and the NAOJ. The Green Bank Observatory is a facility of the National Science Foundation operated under cooperative agreement by Associated Universities, Inc.
\end{acknowledgements}
\bibliography{astrochemistry.bib}{}
\bibliographystyle{aa} 
\end{document}